\documentclass[aps,pra,twocolumn,showpacs,amssymb]{revtex4}
\usepackage{graphicx}

\begin{document}

\title{Black holes and information theory}

\author{Jacob D. Bekenstein}\thanks{\emph{Author's address:} Racah
Institute of Physics, Hebrew University of Jerusalem, Givat Ram,
Jerusalem 91904, Israel}

\begin{abstract} During the past three decades investigators have
unveiled a number of deep connections between physical information and
black holes whose consequences for  ordinary systems go beyond what has
been deduced purely from the axioms of information theory.   After a
self-contained introduction to black hole thermodynamics, we review from
its vantage point topics such as the information conundrum that emerges
from the ability of incipient black holes to radiate, the various entropy
bounds for non-black hole systems (holographic bound, universal entropy
bound, etc) which are most easily derived from black hole
thermodynamics,  Bousso's covariant entropy bound, the holographic
principle of particle physics,  and the subject of channel capacity of
quantum communication channels.
\end{abstract}


\maketitle

\section{\label{sec:intro}   Introduction} 
                   
Black holes entered the stage of natural science---as an astrophysical
paradigm---when British scientist and cleric John Michell~\cite{michell}
and  French mathematician, astronomer and peer P. S. de
Laplace~\cite{laplace} independently remarked that  a star with a
sufficiently large ratio of mass to radius cannot be observed by its own
light because for such configuration  the escape velocity exceeds the speed
of the light corpuscles.  General relativity, the modern gravity theory,
sharpened the definition of `black hole': no longer a mass from which light
cannot issue to large distance, but a region of space  rendered causally
irrelevant to all its environment by gravity in the sense that no signal, by
light, massive particles or whatever, can convey
\emph{information} about its nature and state to regions outside it. 
Information is the key concept here, as emphasised by
Wheeler~\cite{wheeler}.

Nothing just said, however, prepared investigators for the surprising
connections uncovered during the last three decades between the world of
gravity and black holes, and the realms of thermodynamics and
information.  Though figuratively speaking a black hole is a tear in the
fabric of spacetime, albeit one that weighs and moves along very much
like any particle, physics demands that such an entity be endowed with
thermodynamic attributes---entropy, temperature, etc
\cite{NC,PRD1}---as well as quantum properties like the ability, first
uncovered theoretically by Hawking
\cite{hawking1}, to radiate spontaneously in flouting defiance of the
popular definition of `black hole' as an entity incapable of shining on its
own.    

Hawking's radiance engenders a deep problem that has troubled
researchers for over two decades.  The matter which collapses to form a
black hole can be imagined to be in a quantum pure state.  After its
formation the black hole radiates spontaneously, as just mentioned, and
according to calculations, this radiation is in a thermal state, that is a mixed
quantum state.  When this radiation has sapped all mass from the black
hole so that it effectively evaporates, we are left with just a mixed
quantum state of radiation.  That is, the black hole has catalysed
conversion of a pure state into a mixed state, in contradiction to the
principle of unitary quantum evolution.  Mixed means entropic, and so one
can view what has happened as a loss of information.  This is the gist of
the information paradox, which cannot be said to have been settled to
everybody's satisfaction, but which has stimulated thought in gravity
theory both in the gravitation and particle physics camps.  

Black hole entropy has been found to enter into the second law alongside
its more common sibling, matter and radiation entropy.  From the
corresponding generalised second law (GSL)\cite{NC,PRD2}, which has
meanwhile received strong support from a variety of
\emph{gedanken} experiments, it can be inferred that it is physically
impossible to pack arbitrarily large entropy into a region with given
boundary area, or into a given mass with definite extension. These
conclusions (holographic and universal entropy bounds) are quite at
variance with expectations from extrapolation of daily experience (RAM
memories are getting smaller as they shoot up in capacity), as well as with
well understood consequences of quantum field theory, one of the pillars
of contemporary theoretical physics.  It has even been claimed that we
stand at the threshold of a conceptual revolution in physics, one that will
relax the tensions alluded to.  A popular introduction to the matters just
mentioned may be found in Ref.~\cite{SciAm}.  

With one exception, this review focuses on those aspects of physical
information which are not usually discussed by the standard methods of
(quantum) information theory.  Thus after the introduction of black hole
physics (Sec.~\ref{sec:nutshell}) and thermodynamics
(Sec.~\ref{sec:thermo}) and discussion of the information paradox
(Sec.~\ref{sec:paradox}), we review in Sec.~\ref{sec:holo} the holographic
entropy and information bound from the point of view of the GSL, and in
Sec.~\ref{sec:principle} the intimately related holographic principle which
acts as a bridge between information theory, particle physics and
cosmology.  There follows in Sec.~\ref{sec:unibound} an account of the
universal entropy and  information bound, and its origin.  We continue
with  Sec.~\ref{sec:flow}, an essay on the uses of the GSL to set bounds on
quantum channel capacity.  This is a standard topic in information theory
which is enriched by the methods here described.  Sec.~\ref{sec:sum}
summarises the findings.

\section{\label{sec:nutshell}  The black hole in a nutshell} 

One striking thing about the black hole phenomenon, in contrast to other
astrophysically related paradigms, is that it is scale invariant.  A planet is a
planet only if its mass is between a fraction of Earth's and a few times
Jupiter's, and a star shines as such only if its mass is between seven
hundredths and about a hundred solar masses.  Thus, even though
Newton's gravity law has no preferred scale (the only constant involved is
Newton's $G$), planets or stars exist in different mass ranges because of
scales inherent in the matter making them up.  By contrast, classically  a
black hole can have any mass, and its characteristics are the same for all
masses.  This comes about because the laws of relativistic gravity,
Einstein's 1915 field equations, just as Newton's older formulation, do not
involve a preferred scale, \emph{and} because a black hole is
conceptually distinct from any matter (which does have scales) out of
which it originated.    Quantum effects modify the above claims slightly: a
black hole's mass cannot be below a Planck mass ($\sim 2\times 10^{-5}$ 
g) because if it where, the hole would then be smaller than its own
Compton length, and would thus not exhibit the black hole hallmark, the
event horizon.  

The horizon is the boundary in spacetime between the region inaccessible
to distant observers, and the outside world.  It can be crossed only inward
by particles and light.  When viewed at a fixed time (more exactly on a
spacelike surface), it has the \emph{topology} of a two-sphere, namely it
is a closed simply connected 2-D surface.  When a black hole is nearly
stationary it is useful to talk about a typical scale for it, e.g., its radius
$r_{g}$ if it is truly spherical.  The mentioned scale invariance of black hole
physics requires  $r_{g}$ to be proportional to the black hole's mass $m$,
and its average density
$\langle\rho\rangle$ to be proportional to
$m^{-2}$.  For example, in general relativity a spherical electrically neutral
black hole is described by the Schwarzschild solution~\cite{MTW}, a
particularly simple solution of Einstein's equations, which tells us that
\begin{eqnarray} r_{g}&=&{2Gm\over c^{2}}=1.49\times 10^{-13} 
\Big({m\over 10^{15}\textrm{g}}\Big)\,\textrm{cm}
\label{radius}
\\
\langle\rho\rangle&=&{3c^{6}\over 32\pi G^{3}m^{2}}=7.33\times
10^{52}\Big({10^{15}{\rm g}\over
m}\Big)^{2}{\textrm{g}\over\textrm{cm}^{3}}
\label{scale}
\end{eqnarray} The arbitrary mass $10^{15}$ g  that we have introduced to
make the orders of magnitude transparent is of the order of that of a
modest mountain.  It is clear why black holes are popularly regarded as
smallish and hugely dense.  But this is not always true: in our own Milky
Way's core lurks a black hole 20 million kilometers across with an average
density  about that of water.  Of course, black holes originating from
stellar collapse, and those suspected to have survived the rigours of the
early universe, are much smaller and denser.

Black holes cannot shrink; this much seems clear from the fact that they
can devour matter, radiation, etc. but cannot give any of these up.  Penrose
and Floyd~\cite{penrose},  Christodoulou~\cite{christodoulou} and
Hawking~\cite{hawking} independently showed that the said inference is
correct in classical physics if precisely stated: in almost any transformation
of a black hole, its horizon area will increase, and it cannot decrease under
any circumstance.   The `classical' qualifier is critical; Hawking himself
was soon to demonstrate the limitations of this ``area theorem'' once
quantum processes intervene.  But the theorem, though classical in scope,
has turned out to be crucial to developments in black hole physics, not to
mention to astrophysical applications. 
 
Not only are black holes devoid of specific scales; they also lack the wide
variety of individual traits that characterise stars and planets.  A star's
observable aspect, including its spectrum---the stellar
fingerprint---depends very much on its chemical make up.  Stars rich in
the elements heavier than helium have more complicated spectra than do
stars poor in them.    There is a gamut of stellar chemical compositions and
an equally wide range of spectral types.   By contrast, in general relativity
and similar gravity theories, all the black hole solutions describing
stationary charged and rotating black holes form a single three-parameter
family, the Kerr-Newman (KN) black hole solution~\cite{MTW}.

\vspace{0.2in}
\includegraphics[height=2.3in]{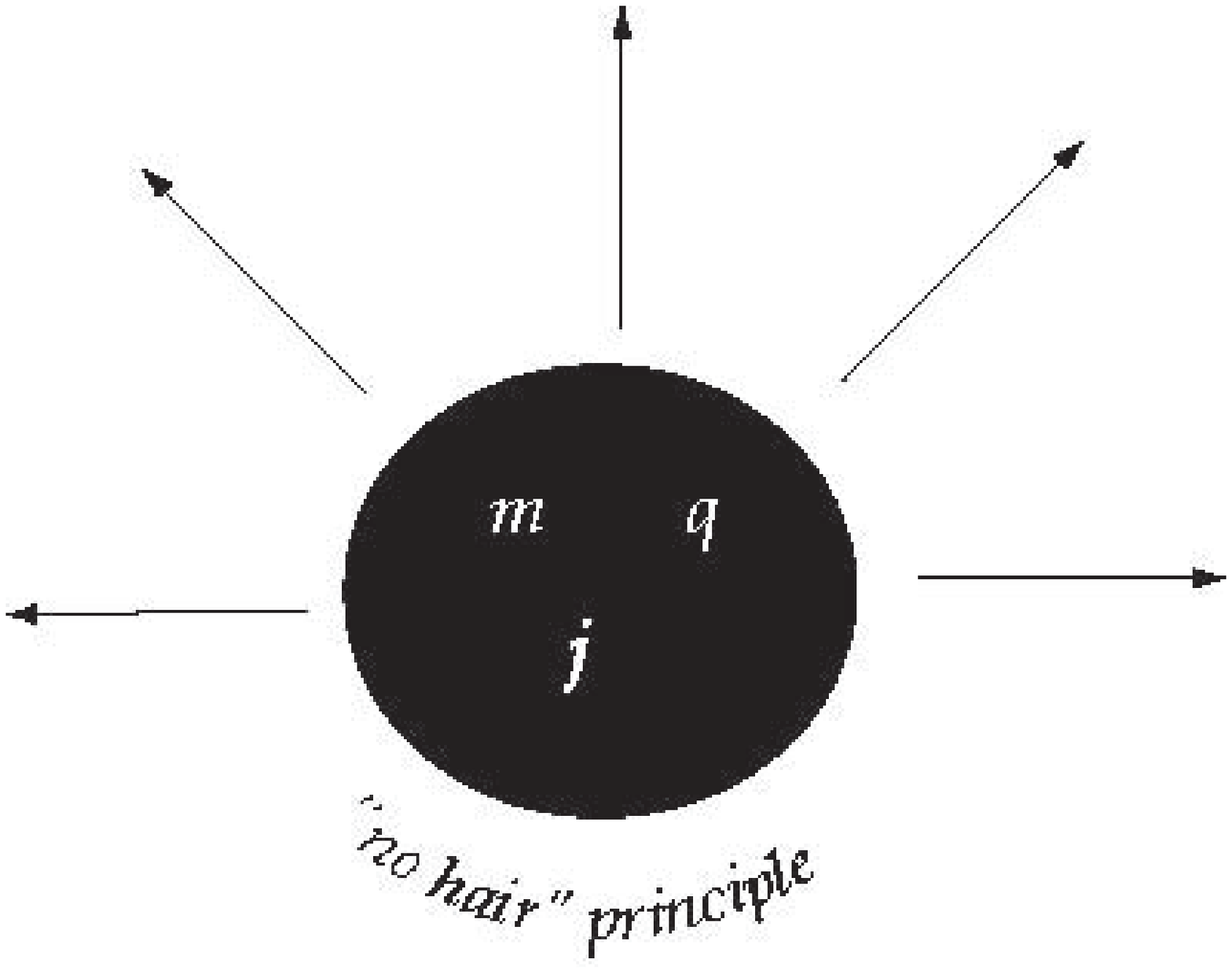}\\
{\bf Fig.~1: A black hole's primary observables---gravitation and
electromagnetic fields, Hawking emission (arrows)---as well a secondary
ones, e.g. shape of the horizon, are entirely determined by its mass $m$,
electric charge $q$ and angular momentum $j$ (this last is here
responsible for the oblateness of the horizon.\hfill}
\vspace{0.2in}

Consequently, all of a black hole's observable traits are widely thought to
depend only on its mass, angular momentum $j$ and electric charge
$q$ (Fig.~1).   Only these three and similar parameters (like
the magnetic monopole,  as yet unobserved in nature) are observables of a
black hole.   Wheeler~\cite{wheeler}, who first emphasised the parsimony
of a black hole's description, coined the maxim ``black holes have no hair'':

\begin{quote}  In standard gravity theory the most general stationary
black hole exterior is described by the KN solution with $m, q$ and $j$ as
its only parameters. 
\end{quote}

It might seem paradoxical to characterise the horizon as blocking all
access to information about the black hole's interior while simultaneously
maintaining that, say, electric charge of a black hole is observable. 
Actually there is no contradiction; the charge of the hole is the charge of
the object that collapsed to make it.  It could always be determined by
Gauss theorem and, by charge conservation, always has the same value.   
Mass and angular momentum are similar in this respect; both can be
determined from particular features of the hole's exterior spacetime
geometry.

\section{\label{sec:thermo}   Black hole thermodynamics}

How can a black hole---a blemish in spacetime---be endowed with
thermodynamics ?  Thermodynamics is a successful description of a
system provided it makes sense to describe the latter by merely a few
parameters: energy, volume, magnetisation, etc., at least at sufficiently
large scale.    Otherwise a more complicated statistical mechanic or kinetic
approach is indicated.  We have seen that a black hole is fully described, as
far as an outside observer is concerned, by just three parameters: $m, q$
and
$j$.  No need to describe the matter that went to form the black hole in all
gory detail.  Hence thermodynamics seems an appropriate paradigm for
black holes.  

What are its variables ? Black hole mass $m$, in the role of energy, is a
typical thermodynamic parameter.  Charged and rotating thermodynamic
systems are, likewise, known, so $q$ and $j$ can be admitted.  But to have
a complete set of observable thermodynamic parameters for a black hole
one still requires entropy.  And it is plain that black hole entropy cannot be
identified with the entropy of matter that went down the black hole, for it
together with the matter becomes unobservable in the course of
collapse.     The fact that horizon area
$A$ tends to increase, and is definitely precluded from decreasing,
suggests it represents the requisite black hole entropy.  Various
\emph{gedanken} experiments together with Wheeler's remark~\cite{PT}
that the Planck length $\ell_{P}\equiv (G\hbar/c^{3})^{1/2}$ (the
Compton length corresponding to Planck's mass)  should play a crucial
role here, motivated me to assert that black hole entropy, $S_{BH}$, is
proportional to  $A/\ell_{P}{}^{2}$~\cite{NC,PRD1}.     Is this reasonable ?
 
The fact that the requisite $S_{BH}$ must be exclusively  a function of
$A$ (also when $q$ and $j$ do not vanish) is most clear from a latter
argument by Gour and Mayo~\cite{gour_mayo,limits}.  For a KN black
hole the area is easily calculated from the metric~\cite{MTW}, namely
\begin{equation}  A=4\pi[(M+\sqrt{M^{2}-Q^{2}-a^{2}})^{2}+a^{2}],
\label{area}
\end{equation} where $M\equiv Gmc^{-2}, Q\equiv\surd G qc^{-2}$ and
$a\equiv jm^{-1}c^{-1}$ are three length scales which completely specify
the black hole ($M$ is just half our previous $r_{g}$ in the Schwarzschild
case $q=j=0$).  The black hole exists only when
$Q^{2}+a^{2}\leq M^{2}$.   We infer from Eq.~(\ref{area}) that
\begin{equation} d(mc^{2})=\Theta dA+\Phi dQ+\Omega dj
\label{first_law}
\end{equation} with
\begin{eqnarray}
\Theta &\equiv& c^{4}(2GA)^{-1}(r_{g}-M)
\\
\Phi &\equiv&q\,r_{g} (r_{g}^2+a^{2})^{-1} 
\\
\Omega &\equiv& j\, M^{-1}(r_{g}^2+a^{2})^{-1}.
\label{derivatives}
\end{eqnarray} 

Since $mc^{2}$ is the energy, Eq.~(\ref{first_law}) has the aspect of the
first law of thermodynamics $TdS=dE-\Phi dQ-\Omega dj$ as applicable
to a mechanical system whose electric potential and rotational angular
frequency are $\Phi$ and $\Omega$, respectively.  Indeed, study of the
motion of charged test particles about a KN black hole shows the above
defined $\Phi$ to be the electric potential at the hole's horizon, while
$\Omega$ is the uniform angular frequency with which infalling particles
are entrained by the horizon---surely a good definition of rotational
frequency of the black hole.  Thus, if a black hole is to have a
thermodynamics (first law at least), we must identify
$\Theta dA\leftrightarrow T_{BH} dS_{BH}$ with $T_{BH}$ the hole's
temperature.  It follows that $S_{BH}=f(A)$; black hole entropy depends
on $m, q$ and $j$ only through the combination $A$.    We then recognise
that $T_{BH}=\Theta/f'(A)$.  Why
$f(A)$ must be linear will be explained shortly.

Entropy lost into black holes cannot be kept track of, and so one should
not, in ordinary circumstances, discuss entropy \emph{inside} black
holes.  Thus the ordinary second law must be given a generalised form.  
Incorporating additivity of all entropy in analogy with other
thermodynamics we get the generalised second law
(GSL)~\cite{NC,PRD2}:
\begin{quote} The sum of black hole entropies together with the ordinary
entropy outside black holes cannot decrease.
\end{quote}~Fig.~2  furnishes an example.  The GSL reduces to
the ordinary second law when black holes are absent, and to the area
theorem if matter and radiation are absent (the last provided $f'(A)> 0$).

Which entropy exactly is covered by the stipulation ``ordinary entropy'' in
the GSL's statement ?  After all, the entropy we associate with some matter
depends on the `resolution' of our description.  If this last is rather coarse
and ignores atoms, then we refer to the chemist's thermodynamic
entropy.  But if we include atomic and subatomic degrees of freedom, then
there may be further contributions to the entropy at sufficiently high
temperatures.  It is easy to see that the ``ordinary entropy'' must be taken
to mean the entropy calculated from statistical mechanics applied to
\emph{all}  degrees of freedom in matter and radiation, no matter how
recondite.  The reason is that the GSL is fundamentally a gravitational law,
and gravitation is aware (via the equivalence principle) of energy residing
in \emph{all} degrees of freedom, no matter how deep they may lie.  For
example, string degrees of freedom should be taken into account if strings
are taken as the fundamental entities. 

\vspace{0.2in}
\includegraphics[height=1.6in]{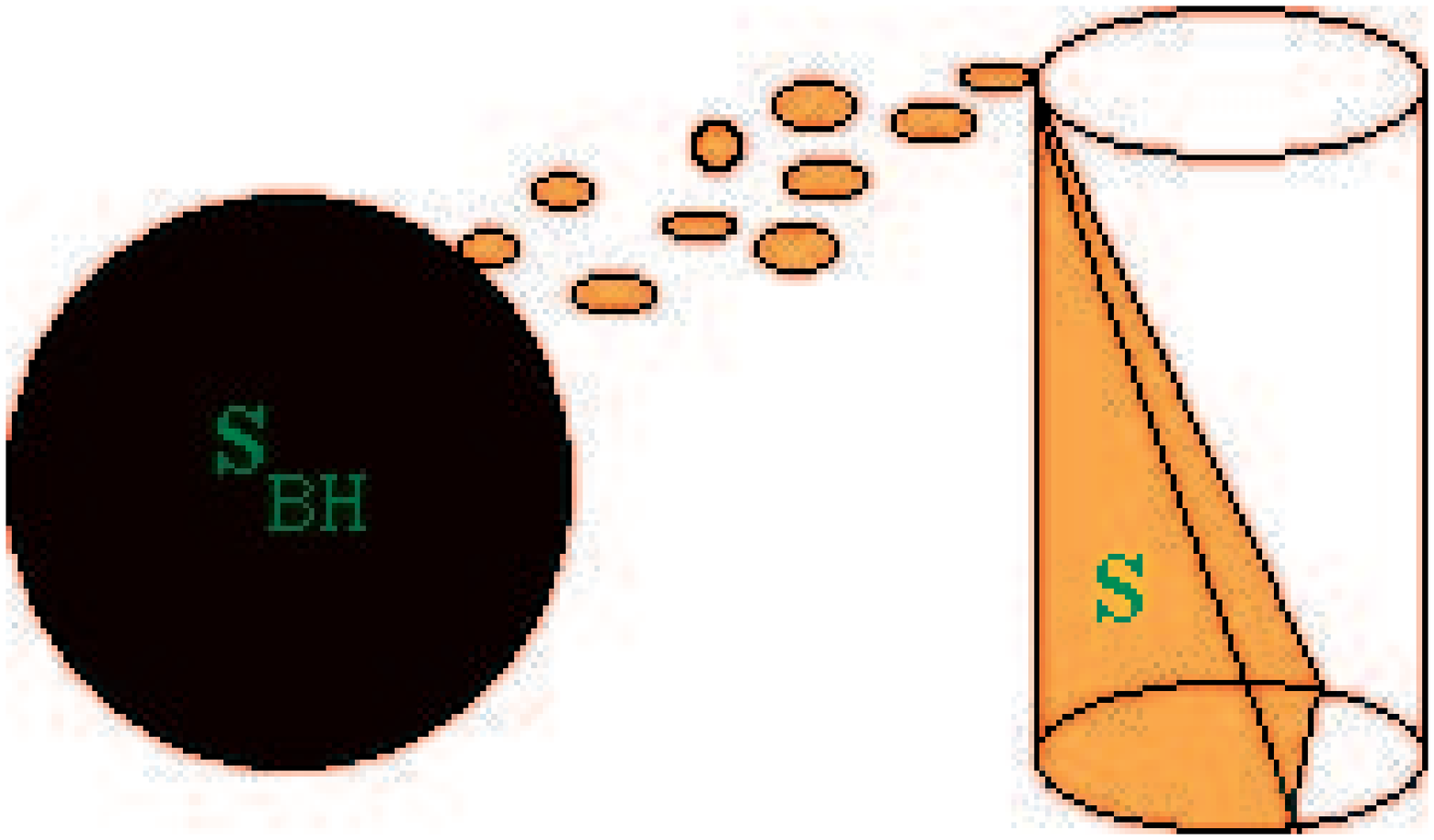}\\
{\bf Fig.~2:  Illustrating the GSL.  The black hole entropy $S_{BH}$ must
increase at least by an amount equal to the entropy of the glassfull of tea
poured down the black hole.\hfill}
\vspace{0.2in}

The GSL makes a linear $f(A)$ seem most reasonable since addition of
areas of black holes is a meaningful procedure (even in general relativity),
and the `area theorem' guarantees that the sum of areas (but not the sums
of other combinations of black hole parameters) will increase.  However,
one might be leery of the linear $f(A)$ because it implies that doubling $m,
q$ and
$j$ of a KN black hole \emph{quadruples} its $S_{BH}$, so that black hole
entropy for one black hole is not extensive as for material systems.  
However, we now show that the seemingly more palatable alternative
$f(A)\propto \surd A$ is excluded together with all laws of the form
$f(A)\propto A^\gamma$ with $\gamma\neq 1$.
 
According to the area theorem, when a capsule containing some matter is
dropped into a black hole, the latter's horizon area must increase.  But is
the consequent growth in $S_{BH}$ sufficient to compensate for the loss of
the capsule's entropy, $S_{cap}$, as the GSL demands ?  One can try
pushing the GSL against a wall by inserting the capsule (with rest mass
$\mu$ and radius $b$) in a most gentle manner so as to minimise the area
increase
$\Delta A$.   Purely mechanical arguments~\cite{PRD1,brazil} show that
for a generic KN black hole sufficiently big to accept the capsule and much
more massive than it, there is a lower bound $\Delta A\geq 8\pi G\mu b
c^{-2}$ strikingly independent of the black hole parameters.  It
immediately follows that $\Delta S_{BH} \geq f'(A) 8\pi G\mu b
c^{-2}\geq S_{cap}$.  The second inequality comes from the GSL since in
the infall the capsule's entropy is lost.  Were we to choose $f(A)\propto
A^\gamma$ with
$\gamma$ constant, we would obviously be faced with a violation of the
GSL for large $A$ if $\gamma<1$, or for small $A$ if $\gamma>1$.  
 
 We can conclude that $f(A)$ cannot be an exact power law, except for the
trivial one with $\gamma=1$.  Thus we adopt
 \begin{equation} S_{BH}=\eta A/\ell_{P}{}^{2}
 \label{S}
\end{equation} with $\eta$ a constant; $\ell_{P}$ is introduced for
dimensional reasons. Our earlier result $T_{BH}=\Theta/f'(A)$ thus gives
(except where explicitly stated otherwise, temperature is in units of
energy) 
\begin{equation} T_{BH}= (c\hbar/2\eta A)\sqrt{M^{2}-Q^{2}-a^{2}}.
\label{T}
\end{equation}  The smaller the hole, the hotter it is (see
Eq.~(\ref{Tnum})).   Black holes which have about as much angular
momentum (or charge) as permitted are especially cool.   Finding $\eta$ is
obviously the next logical step. 

Attempts to understand the simple formula for black hole entropy from
more fundamental points of view are legion.    Bombelli, Kaul, Lee and
Sorkin~\cite{BKLS}, and later and independently Srednicki~\cite{sred}
gave reasons to believe that black hole entropy is related to entanglement
entropy arising from the tracing out of those degrees of freedom that are
localised beyond the horizon.  Early attempts of Thorne and
Zurek~\cite{TZ} and independently 't Hooft~\cite{thooft2} sought to
identify black hole entropy with the entropy of the thermal radiative
``atmosphere'' of the black hole (see Sec.~\ref{sec:paradox}).     We may
also mention here some of the many attempts to relate it to the degrees of
freedom of strings associated with the black
hole~\cite{strom_wafa,malda_strom}, or to  the quantum gravity degrees
of freedom of the horizon, be it in the context of conformal field
theory~\cite{carlip,solo}, of loop quantum gravity theory~\cite{ash}, or of
more heuristic schemes~\cite{mukh,MG8,brazil}.

\section{\label{sec:paradox}   Hawking radiation and the information
paradox}

 In the everyday world hot objects radiate.  In 1974 Hawking demonstrated
theoretically that a black hole formed by collapse does
likewise~\cite{hawking1}.  In essence his quantum field theoretical
calculation performed on a prescribed classical gravitational background
shows that if a quantum field is in the vacuum state in the presence of an
object which begins to collapse to a black hole, then as the object's radius
nears the horizon's, the state of the field in the hole's exterior approaches
that of  thermally distributed radiation with a temperature of the form
(\ref{T}) with
$\eta={\scriptstyle 1/ \scriptstyle 4 }$ .  This temperature is the same
whatever the field, e.g. scalar, electromagnetic, neutrino, etc.   Thus
Hawking's result calibrated the black hole entropy and temperature
formulae  (\ref{S}) and  (\ref{T}):
\begin{eqnarray} S_{BH}&=&2.65\times 10^{40} (m/10^{15}{\rm g})^{2}
h_{1}
\label{Snum}
\\ T_{BH}&=&1.23\times 10^{11} (10^{15}{\rm g}/m) h_{2}{}^{0}{\rm K}
\label{Tnum}
\end{eqnarray} where $h_{1}(Q/M,a/M)$ and $h_{2}(Q/M,a/M) $ are
two known dimensionless functions of order unity, both exactly equal to
unity for
$Q=a=0$.   

For comparison, the sun (mass $2\times 10^{33}$ g) has an entropy of
order $10^{58}$ and central temperature $1.6\times 10^{7}{}^{o}{\rm K}$.
On the astronomical scale black holes are thus very entropic and cool.  It is
consistent with the GSL that a solar mass black hole have an entropy
larger than that of a solar mass star which might have been its
predecessor.  But why should the holes's entropy be the larger by many
orders of magnitude ?  Boltzmann's principle that a system's entropy is the
logarithm of the number of microscopic configurations compatible with
that system's macroscopic properties, together with the ``no hair''
principle, suggests that black hole entropy is large because a black hole's
aspect cannot tell us precisely which type of system gave rise to it.  This
extra lack of ``composition information'' over and above that about
specific microscopic configurations may be what makes black hole entropy
large.  A black hole stands for a large amount of missing information.

Hawking noticed a conundrum when black hole radiation is considered in
light of the unitarity principle of quantum theory~\cite{hawking2}.  One
can imagine a black hole formed from matter in a pure state, e.g. a
gravitating sphere of superfluid at $T=0 ^{0}$K.  The unitarity principle
would thus require that the system always remain in a pure state. 
 The fact that a black hole with large entropy forms is not in itself the real
problem.  Although nonzero entropy is a property of a mixed state, it is an
everyday sight sanctioned by the second law that entropy  can just appear
when there was none before.  This is understood as reflecting classical
coarse graining or tracing out of some quantum degrees of freedom in a
fundamentally pure state, both transpiring for operational reasons.  The
conundrum arises only in the aftermath of the black hole (and,
incidentally, is unrelated to the oft made observation that according to
\emph{some} observers the black hole horizon never quite forms).  
 
 Hawking's radiation drains the hole's mass (also its angular momentum)
on a finite time scale.  Using the Stefan-Boltzmann radiation law $P =(4\pi
R^{2}) \sigma T^{4}$ and Eqs.~(\ref{Tnum}) and (\ref{radius}), we
estimate, c.f. Eq.~(\ref{flux}) below, for each species of quanta radiated by
a Schwarzschild black hole
 \begin{equation}  {dm\over dt}\approx -4.02\times
10^{-6}\Big({10^{15}{\rm g}\over m}\Big)^{2} {\rm g\, s}^{-1},
 \label{dm/dt}
\end{equation}  with an order of magnitude correction coming from the
gravitational redshift, general relativistic geometrical factors, and particle
statistics (Bose or Fermi). Obviously as the black hole looses mass, it
radiates faster and the mass loss accelerates.   Calculations give no hint
that the evaporation can be arrested before $m$ descends to Planck  mass
scale, by which time one is dealing with a pure quantum gravity
phenomenon.  Evaporation of the black hole to nothing, or at the very
least to a Planck scale object, must thus take less than
$10^{20}(m/10^{15} {\rm g})^{3} s$.  A black hole with a radius a little
smaller than the proton's (see Eq.~(\ref{radius})) can thus have a lifetime
briefer than our universe's age.   The evaporation is thus slow but not an
hypothetical phenomenon.
 
 Hawking's original calculation~\cite{hawking1} and many others since
then showed that the radiation is thermal, both in its Planck-like spectrum
and in the lack of correlations between different radiation modes.  It thus
seems that a pure state can be converted into a mixed one through the
catalysing influence of a black hole!  Hawking~\cite{hawking2} was led
by this to assert that gravity violates the unitarity principle of quantum
theory. This means the mixed character of the final state is dictated by
physics, and is not the result of the way we choose to describe the system. 
Since the final state has a lot of entropy (of order of the intermediate black
hole's entropy by the GSL), we are faced with a large intrinsic loss of
information.  To be sure, Hawking's inference has remained controversial:
whereas general relativity investigators have tended to accept this
conclusion, particle physicist have stood by the unitary principle and
orthodox quantum theory.  Many resolutions have been offered.  In order
to categorise them it is useful to draw an analogy between our problem
and the following ``experiment'' attributed to S. Coleman.
 
A cold piece of coal is illuminated by a laser beam.  The system is in a pure
state: coal in its ground state and beam in a coherent state (analogous to
the sphere of superfluid).  Experience tells us the coal will heat up and
radiate (black hole forming and radiating).  The beam is interrupted (no
matter is thrown into black hole after its formation).  The coal cools while
radiating thermally (Hawking radiation).  The coal cools totally  and
returns to its ground state (black hole evaporates), which is, of course,
pure.  In both cases we are left with a mixed state---thermal radiation is as
mixed as can be.  Nobody doubts that unitary is respected in the coal-laser
system.  The information conundrum in that case is unravelled by the
remark that subtle correlations between the early and late radiation take
care to preserve the purity of the state once the correlation with the coal is
broken by its reaching ground state.  Any coarse inspection of the
radiation, which would necessarily be local, would, by virtue of the
implicit tracing out, reveal a mixed thermal state.

Many have argued by analogy that subtle correlations in the actual
Hawking radiation preserve its overall pure state status after the black
hole is gone.  The fact that Hawking's and like calculations reveal no such
correlation is thought to be due to their semiclassical character  which
ignores the quantum degrees of freedom of gravity.  Certain model
quantum gravity calculations have supported this point of view.  

A different approach to resolving the paradox posits that a Planck scale
remnant is always left after Hawking evaporation, and that it escapes the
fate of total evaporation by means of quantum gravity modifications to
Eq.~(\ref{dm/dt}).  It is impossible, in the present state of the art of
quantum gravity research, to check this possibility thoroughly.  It does
involve belief in objects of dimension $10^{-33}$ cm whose information
content corresponds \emph{at least} to the entropy
$10^{40}$ characteristic of a black hole sufficiently light to evaporate to
Planck scale in the lifetime of our universe (see Eq.~(\ref{Snum}) and
conclusions stemming from (\ref{dm/dt})).   But this is
problematic~\cite{bek94}; as we shall see, such large information content
in such small size conflicts with recent ideas.
  
 A third way out of the information paradox which respects unitarity, is
the supposition that a black hole always gives rise to another universe
which may be reached, in principle, through the black hole.  Certain black
hole solutions of Einstein's equations, like Reissner and Nordstr\"om's
one describing a nonrotating charged black hole~\cite{MTW}, do show
such a universe connected to the black hole's interior and lying to the
future of the universe in which the black hole formed.  It is not clear
whether the later universe will indeed appear in the aftermath of realistic
collapse.  But if this point is granted, then the idea is that unitary evolution
proceeds as always into the new universe so that no loss of information
actually occurs from a universal point of view.  The local observer in the
universe where the black hole formed does retain the impression that
information has been lost.
 
 In summary, there may be resolutions to the information paradox.  But
the preservation of information is only manifest at an ideal level of
scrutiny (being aware of all universes, analysing radiation emitted
millions of years ago jointly with fresh one).  For \emph{practical
purposes}, information does disappear in the presence of black holes.
 
 \section{\label{sec:holo}   The holographic bound}
 
 The GSL  immediately suggests the existence of information (or entropy)
bounds for non-black hole objects.  The first such derived in this way, the
universal entropy bound~\cite{bound}, will be reviewed in the next
section.  Here we take up the holographic bound which is in many ways
easier to comprehend.
 
 L. Susskind~\cite{susskind} proposed the following \emph{gedanken}
experiment.  Take a neutral nonrotating spherical object containing
entropy $\mathcal{S}$  which fits entirely inside a spherical surface of area
$\mathcal{A}$, and allow it to collapse to a black hole, which by
symmetry must be of the Schwarzschild type.   Evidently the black hole's
horizon area is smaller than $\mathcal{A}$, but by the GSL $S_{BH}$ must
exceed $\mathcal{S}$.  It follows that 
 \begin{equation}
\mathcal{S}\leq {\mathcal{A}\over 4 \ell_P{}^{2}}.
\label{holo}
 \end{equation}
 We have included the equality in order that a black hole itself partake of
this, the holographic bound.  
 
 By the connection between entropy and information, bound (\ref{holo})
implies a generic ``holographic'' bound on the information inscribed in
any isolated object, a bound which can be stated exclusively in terms of
the area of some bounding surface (see Fig.~3).  Like the
holographic bound on entropy, this second bound is counterintuitive. 
Has it not been obvious for generations that, other things being equal,
information capacity scales with volume of the information registering
\emph{milieu} ?  But if so, as the scale of the system goes up, the growth
in volume must outstrip the growth of area bringing about a conflict with
the assertion of the holographic bound.  The resolution to the quandary is
that before the crossover point is reached, the information storage system
has already collapsed to a black hole which, of course, cannot be used to
store information  useful to external observers~\cite{SciAm}. 
 
\vspace{0.2in}
\includegraphics[height=2.0in]{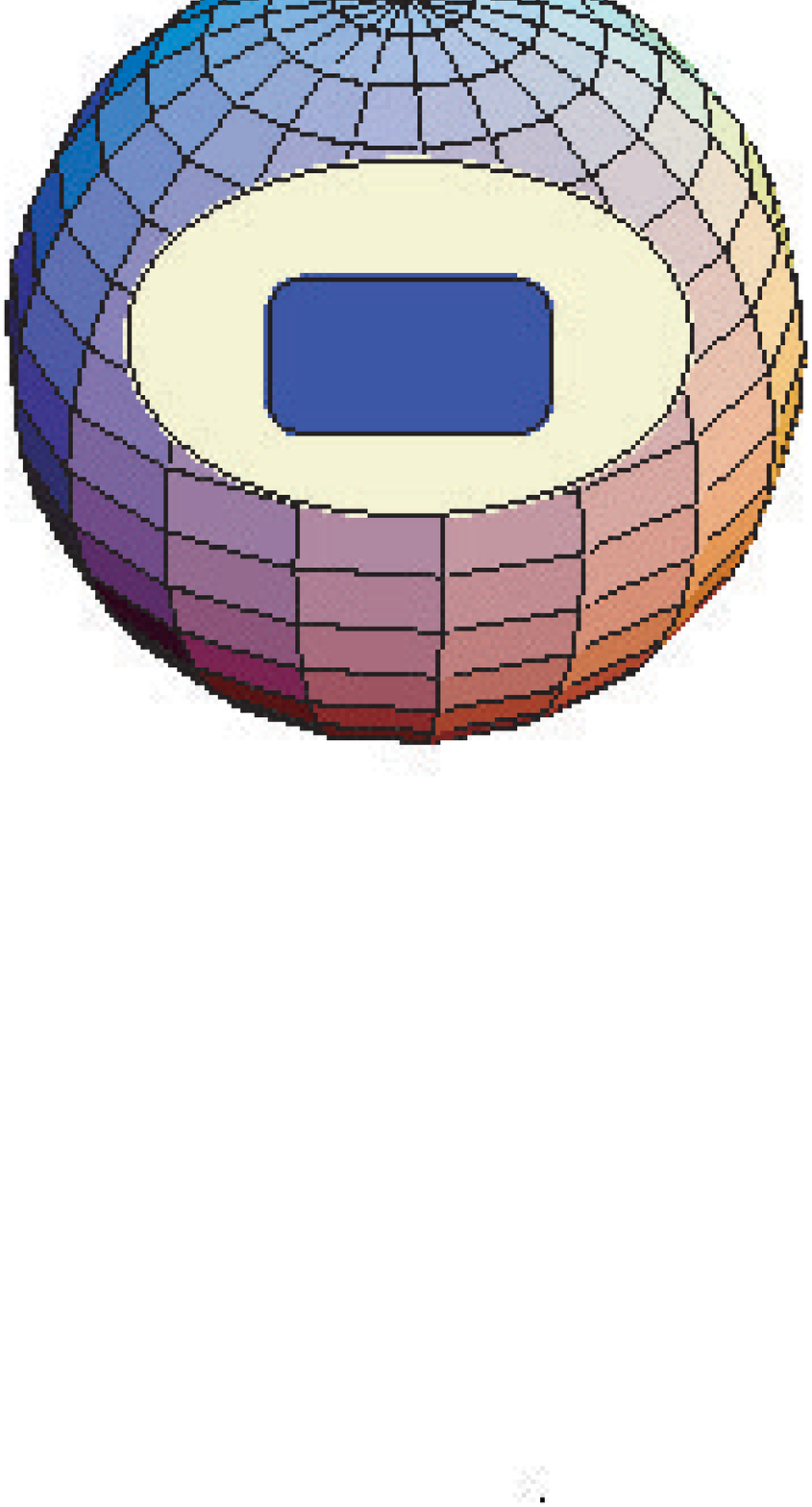}\\
{\bf Fig.~3: Illustrating Susskind's version of the holographic bound. 
The information capacity in nits of an information storage of arbitrary
construction (purple object) is bounded from above by a quarter of the
area of an enclosing spherical (or other closed) surface measured in
squared Planck lengths.\hfill}
\vspace{0.2in}

 But how general is bound (\ref{holo}) ?  It might appear that the bound
should not be used for an object which cannot spontaneously collapse, e.g.
the Earth.  But, at least in two kinds of circumstances, its applicability is
justifiable even in this situation.  If the system in question is weakly
self-gravitating (as are most everyday, laboratory  and astronomical
systems),  namely its mass $\mu$ and radius $R$ satisfy
$G\mu c^{-2}\ll R$, Susskind's spherical collapse can be supplanted by
the infall of the object into an already existing much larger and heavier
black hole (see end of Sec.~\ref{sec:unibound}).   Alternatively, for a
strongly self-gravitating composite system ($G\mu c^{-2}\sim R$), which
is at the same time much larger and more massive than an elementary
particle, e.g. a neutron star, a tiny preexisting black  hole can be used to
catalyse its eventual collapse, with the hole contributing little to the
bookkeeping~\cite{catalyse}.  In both of the above cases the GSL can be
used to recover Susskind's holographic bound (\ref{holo}).  

 Susskind's argument will also apply to a charged object provided it
indeed collapses to a black hole in spite of the Coulomb repulsion.  Due to
the quirks of relativity it is less clear if collapse to a rotating (Kerr) black
hole of a very compact rotating object necessarily involves a contraction of
the bounding area, though this seems likely.  Therefore, the holographic
bound's range of applicability is broad.   It is, however, inappropriate to
apply bound (\ref{holo}) to a system which is not well isolated from its
surroundings.  Gravitational collapse is not controllable, and cannot be
expect to affect the system while exempting surrounding objects from like
fate.
 
 All the above is not to say that the holographic bound, as just stated, is
flawless.  For instance, it fails if applied to our whole universe, particularly
if the latter is infinite, as suggested by contemporary cosmological data.  In
the standard cosmological model the universe contains entropy
(principally of radiation) with some uniform density.  A sufficiently large
sphere can thus contain more entropy than allowed by (\ref{holo}) if
$\mathcal{A}$ is interpreted as its bounding area, because then
$\mathcal{A}$ scales only as the square of the radius of the said volume. 
And if the universe is fairly uniform, the above mentioned crossover point
is not accompanied by collapse.    Likewise, a spherical system already
inside the black hole of it own making will eventually violate (\ref{holo}),
because as it inexorably contracts (as it must by general relativity), its
bounding area  eventually shrinks to zero while, by the ordinary second
law, the entropy it contains cannot decrease.   (Note that we here take the
view of the \emph{interior}  observer). 
 
 R. Bousso~\cite{bousso00,boussorev} introduced a reinterpretation of
formula (\ref{holo}), the covariant entropy bound, which makes it more
broadly valid.  We shall not go into technical details.  Suffice it to say that
$\mathcal{A}$ now refers to the area of any 2-D surface, closed or open,
which satisfies mild technical restrictions.  The
$\mathcal{S}$ refers to all the entropy that ``gets illuminated'' by a
hypothetical brief flash of light emitted perpendicularly from one side of
the surface, with entropy beyond the point where light rays start crossing
not counted.   In cases where the original holographic bound applies, it
can be derived from Bousso's~\cite{bousso00}.  As of this writing,
Bousso's bound has navigated successfully a number of classical
hurdles~\cite{boussorev}. It is, however, known that quantum radiation,
e.g. Hawking's, can cause the Bousso bound's failure~\cite{loewe}.  A 
generalisation of the bound to extend its validity to this situation---but still
short of quantum gravity---has been proposed by A. Strominger and D.
Thompson~\cite{strom}.
 
 \section{\label{sec:principle}     The holographic principle}
 
 Much of the contemporary status of the holographic bound, in whatever
formulation, is due to its intimate connection with G. 't Hooft's
holographic \emph{principle}~\cite{thooft} (in fact the adjective
`holographic's was applied by 't Hooft at the outset to the principle).  
Many workers regard the holographic principle as providing guidelines
for the final theory of nature. The holographic principle asserts that
physical processes in a universe of $\mathcal{D}$ spacetime dimensions
as described by some physical theory, e.g. string theory or a field theory,
are reflected in processes taking place on the $\mathcal{D}-1$
dimensional boundary of that universe (provided it has one) which are
described by a different physical theory formulated in $\mathcal{D}-1$
dimensions.  There is an equivalence between theories of different sorts
written in spacetimes of different dimensions~\cite{maldacena,witten}.  
 
 A concrete example is the equivalence of string theory operating in 5-D
anti-deSitter spacetime and a conformal field theory operating in the 4-D
flat spacetime which constitutes 5-D anti-deSitter spacetime's boundary.  
DeSitter spacetime is a solution of Einstein's equations, as augmented by a
positive cosmological constant, representing a highly symmetric (and
formally empty) universe.  There is much astronomical evidence that our
universe may be headed for a deSitter like phase.  Anti-deSitter spacetime
is obtained from deSitter's solution by switching the sign of the
cosmological constant.  A consequence of the above mentioned
equivalence is the correspondence between properties of a black hole, a
still mysterious entity here conceived in string theory terms, which resides
in the anti deSitter universe and those of black body radiation (by now a
trite subject) in the flat spacetime~\cite{witten}.  Correspondences of this
sort have been used to simplify difficult calculations.  And the equivalence
between the laws is undeniably of philosophical import.  Thus far it has
not proved possible to set up a like holographic correspondence involving
deSitter's universe.
 
 The relation between  holographic \emph{principle} and \emph{bound}
is an informational one.  If processes in the bulk spacetime can be
understood by correspondence with processes on its boundary, then in
some sense the measure of information about the bulk is not so large that it
cannot be bounded in terms of the extent of the boundary, which is the
natural measure for information therein.   For systems in
three-dimensional space, this suggests an information content that scales
no faster than the area of the boundary of the space, as in the holographic
bound.  There is also contrast between bound and principle.  The
holographic bound  is  applicable also to part of the space and the
corresponding boundary (provided, as mentioned above, that the system
in question is truly isolated).  This is in stark contrast to the full
holographic principle which only asserts detailed equivalence of the
physics in two different `universes'.   
 
Although in the public's mind  the holographic principle is associated with
string theory, the two actually stand in conflict, just as do the principle
and quantum field theory.  Fields are continuous, and they live in a
continuum spacetime.  As a result a field has an infinity of possible
different (orthogonal) states in a typical volume (say one whose boundary
is not unduly convoluted), and certainly in a whole universe.  But it is
already clear from the holographic bound that to the given volume can be
associated only a finite entropy, \emph{ergo} a finite number of states. 
Likewise, a string, quite different from a field in other respects, also has an
infinite number of possible states (think of the number of distinct
vibrations of a taut cord).  Since the string can be confined to a given
volume,  this conclusion clashes with the holographic bound. 

 The clash is not confined to finite systems.    The deSitter `universe' has
\emph{infinite} volume, much of it hidden behind an event horizon very
like that of a black hole in many respects, but encompassing the whole
`sky' of the observer.   This horizon has a finite area (whose size is set by
the value of the cosmological constant---a parameter of the physics).  Since
Gibbons and Hawking's early paper on thermodynamics of deSitter
spacetime~\cite{gibb}, entropy has been ascribed to deSitter's horizon
according to the usual black hole rule (\ref{S}), and the GSL is known to
apply~\cite{davies}.    It follows that the entropy of matter (or radiation)
hidden behind the horizon is always finite, even though the universe is
infinite.  This obviously raises challenges for string theory as much as for
field theory. 
    
 \section{\label{sec:unibound}   The universal information bound}
 
The holographic information \emph{bound} is simple;  it is also extremely
lax.   For example, an object the size of a music compact disk would be
allowed by the bound an information capacity of up to $10^{68}$ bits. 
Present technology can only store $10^{10}$ bits on it, and is expected to
improve only by a few orders of magnitude.  It is clear from this and other
examples that the holographic bound, important though it be for matters
of principle,  is of no great practical use.  Can an alternative do better
while still being generally correct ?  Indeed, the hoary universal entropy
bound~\cite{bound} does much better than the holographic bound.  The
original argument for it involves fine points of general relativity; it has
also been attacked on the grounds that it does not properly account for the
phenomenon of quantum buoyancy~\cite{buoy}.  Therefore, we provide
here a much simplified approach to the said entropy bound~\cite{erice}.

\vspace{0.2in}
\includegraphics[height=1.6in]{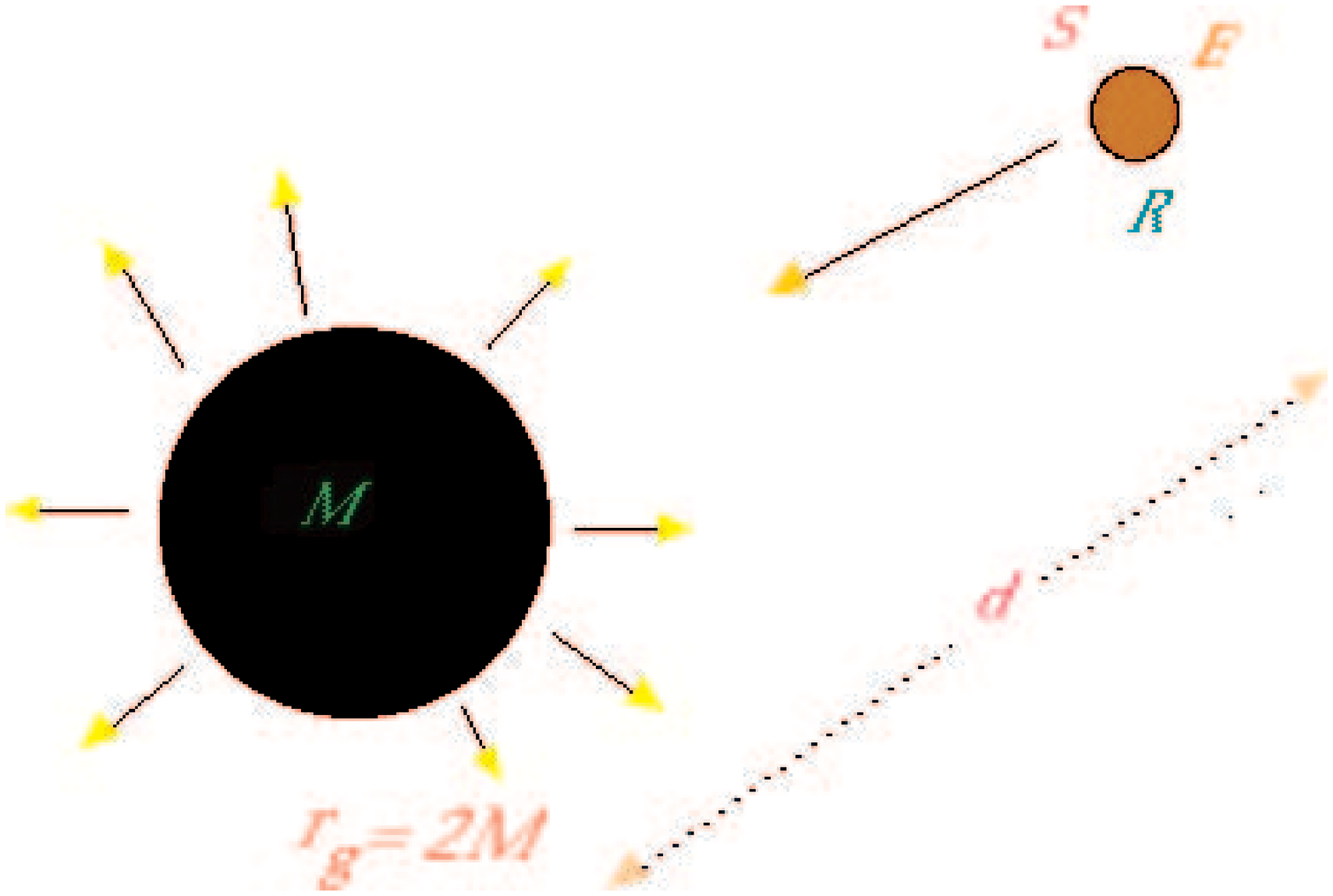}\\
{\bf Fig.~4: Free infall of a macroscopic object into a Schwarzschild
black hole can be used to derive a weak version of the universal entropy
bound.\hfill}
\vspace{0.2in}

Consider the following \emph{gedanken} experiment. Drop a composite 
system ${\cal U}$ (not an elementary particle) of radius
$R$, total energy $E$ (its rest energy) and entropy $\mathcal{S}$ into a
Schwarzschild black hole of mass
$m\gg E c^{-2}$ from a  large distance $d\gg M$ away; $d$ is so chosen
that the Hawking radiance carries away energy  equal to $E$ while ${\cal
U}$ is falling to the horizon where it is effectively assimilated by the black
hole.  This is depicted in Fig.~4.   Upon completion of the
process the black hole mass is back to $m$ and its entropy has not
changed.  Were the emission reversible, the radiated entropy  would be
$E/T_{BH}$ with $T_{BH}\equiv c\hbar(8\pi M)^{-1}$ (see
Eq.~(\ref{T})).  The curvature of spacetime makes the entropy emitted a
factor $\nu$ larger; typical values, depending on particle species, are
$\nu=1.35$--$1.64$ \cite{page2}.  Thus the overall change in world
entropy is 
\begin{equation}
\Delta S =  \nu E/T_{BH} -\mathcal{S}.
\label{netchange} 
\end{equation}

One can certainly  choose  $M=Gmc^{-2}$ larger than
$R$, say, by an order of magnitude so that the system will fall into the hole
without being torn up: $M=\zeta R$ with $\zeta$ of the order a few. Thus
by the GSL (the black hole is unchanged) we obtain
\begin{equation}  
\mathcal{S} < 8\pi\nu\zeta RE/c\hbar. 
\label{bound}   
\end{equation}  This bound applies to an arbitrary composite system. 
This means we must require $R\gg  c\hbar/E $ ($\mathcal{U}$ much
larger than its own Compton length; even a nucleon qualifies). 
Additionally, in the derivation ${\cal U}$ is not allowed to be strongly
gravitating (which would entail $GEc^{-4}\sim R$) because then $m$
could not be large compared to $E$, as we have assumed, if we really
insist that $\zeta$ is of order a few.  We thus have to assume in addition
$GEc^{-4}\ll R$.  Note, however, that the resulting bound on entropy,
(\ref{bound}),  is $G$ independent; gravity was central in the derivation,
but has been swept under the rug in the result.

Note also that it is impossible to infer (\ref{bound}) by using a plain heat
reservoir in lieu of a black hole.  A reservoir which has gained energy
$E$ upon ${\cal U}$'s assimilation, and has returned to its initial energy by
radiating, does not necessarily return to its initial entropy, certainly not
until ${\cal U}$ equilibrates with the rest of the reservoir.  But a
(nonrotating uncharged) black hole whose mass has not changed overall,
retains its original entropy because that depends only on its mass (any
equilibration here is on a dynamical scale, and thus extremely rapid).  In
addition, for the black hole mass and radius are related in a simple way;
this allowed us to replace $T_{BH}$ in terms of
$R$. By contrast, for a generic reservoir, size is not simply related to
temperature.  So black holes are crucial in obtaining (\ref{bound}) without
delving into the thermodynamics of $\mathcal{U}$.

The above derivation cavalierly ignores the effect of Hawking radiation
pressure.   Could it blow ${\cal U}$ outwards ?   We can get an idea of the
effect by calculating the Hawking radiation flux $\mathcal{F}$ via
Stefan-Boltzmann's law as applicable to a sphere with temperature
$T_{BH}$ and radius $2M$.  At distance $r$ from the hole
\begin{equation} 
\mathcal{F}(r)= {{c^2\bar\Gamma\mathcal{N}}\hbar\over 61,440(\pi M
r)^2}.\label{flux}
\end{equation}  Here ${\mathcal{N}}$, a natural constant, is the effective
number of massless species radiated (photons contribute 1 to
${\mathcal{N}}$ and each neutrino species $7/16$), and $\bar\Gamma$
corrects for general relativistic effects, including the fact that the radiating
area is actually a bit larger than $4\pi (2M)^{2}$:
$\bar\Gamma\sim 2$~\cite{page1}.   This energy (and momentum) flux
results in a radiation pressure force $f_{\rm r}(r)=\pi R^2
c^{-1}\mathcal{F}(r)$ on ${\cal U}$.  More precisely, species which reflect
well off ${\cal U}$ are approximately twice as effective at exerting force as
just stated, while those (neutrinos and gravitons) which go right through
${\cal U}$ contribute very little.  

Since the Newtonian gravitational force on ${\cal U}$ is  $f_{\rm
g}(r)=GmEc^{-2} r^{-2}$,
\begin{equation}  {f_{\rm r}(r)\over f_{\rm g}(r)}=
{c\bar\Gamma{\mathcal{N}}_{\rm eff}\,\hbar R^2\over 61,440  \pi 
M^3 E}.
\label{ratio}
\end{equation}  We write here an effective number of species,
${\mathcal{N}}_{\rm eff}$, because, as mentioned, some species just pass
through ${\cal U}$ without exerting force on it.  In addition, only those
radiation species actually represented in the radiation flowing out during
${\cal U}$'s infall have a chance to exert forces.  An Hawking quantum,
just as any quantum in thermal radiation, bears an energy of order
$T_{BH}$, so the number of quanta radiated together with energy $E$ is
approximately  $8\pi M E/c\hbar$.  Our assumption that 
$\mathcal{U}$ is composite ($R\gg  c\hbar/E $) and our stipulation that
$M=\zeta R>R$ together make this number large compared to unity. 
Since a species can exert pressure only if it is represented by at least one
quantum, one obviously has
${\mathcal{N}}_{\rm eff} < 8\pi ME/c\hbar$.  Therefore,
\begin{equation}    {f_{r}(r)\over f_{g}(r)}< {\bar\Gamma R^2\over 7680
M^2}= {\bar\Gamma\over 7680 \zeta^2 }
\ll 1 
\label{newratio}
\end{equation}  Radiation pressure is thus negligible.

We must still check our tacit assumption that $d\gg M$, which, by making
most of the infall take place in the Newtonian regime, exempts us from
having to deal with general relativistic corrections.  We recall that $d$
must be such that the infall time equals the time $t$ for the hole to radiate
energy $E$.  Newtonially speaking,  the time $t$ for free fall of a test body,
from $d$ to $2M$ in the field of a mass $m=c^2 MG^{-1}$ is given
implicitly by $d\approx 2(c^2 t^2 M/\pi^2)^{1/3}$, while
Eq.~(\ref{flux}) gives the estimate $t\approx 7680\pi EM^2 
c^{-2}\hbar^{-1}{\mathcal{N}}^{-1}$ (we have taken
$\tilde\Gamma\approx 2$ and  ${\mathcal{N}}$ as the full species
number).  From these equations and $M=\zeta R$ we get that 
\begin{equation} d\approx 780 (\zeta ER/{\mathcal{N}}c\hbar)^{2/3}M.
\end{equation}  Thus for ${\mathcal{N}}<10^2$ (conservative estimate of
{\it our\/} world's massless particle content) and taking into account
$R\gg  c\hbar/E $, we have $d\gg  36 \zeta^{2/3} M$ for all weakly
gravitating composite systems ${\cal U}$.  For all these we have thus
justified the entropy bound (\ref{bound}).  

How big is the factor $\nu\zeta$ ?  It seems safe to assert that
$4\nu\zeta<10^2$.  Indeed, the original argument~\cite{bound} gave
$2\pi$ for the numerical coefficient in the bound; with this choice it is
referred to as the \emph{universal entropy bound}.   A pleasant spinoff of
the choice $2\pi$ is that the bound then formally applies also to black
holes if we identify $E$ with $m c^2$ and $R$ with $(A/4\pi)^{1/2}$
(with $A$ defined by Eq.~(\ref{area})).  The Schwarzschild black hole is
the only one to saturate the entropy bound~\cite{bound}.  

In our derivation  of  bound (\ref{bound}) $R$ evidently stands for the
\emph{largest} radius of the system $\mathcal{U}$.  It has been
claimed~\cite{bousso03,BFM} that $R$ in the universal bound can
actually be  interpreted as a smaller dimension, if such is available. The
derivation of this improved bound relies on a generalised form of the
purely classical Bousso bound.  The validity of this generalisation has,
however, lately been cast in doubt by a counterexample of V.
Husain~\cite{husain}.

This is also the point at which to notice that the condition of weak
self--gravity, $GEc^{-4}\ll R$ allows us to infer immediately that the
holographic bound (\ref{holo}) is satisfied with plenty of room to spare
just as a consequence of the universal entropy bound. In other words, for
weakly self--gravitating systems (which include most systems known), the
universal entropy bound is much tighter than the holographic one.
 
Being a statement about entropy of a system, the universal bound can also
be derived directly from statistical mechanics for simple (quantum)
systems~\cite{bek84,bek_schiff}.   None of these analytical or numerical
arguments have the simplicity or broad applicability of the argument
described above.  But whatever its derivation, the universal bound
automatically provides us with a bound on information. In words, the
ceiling on the capacity in bits is of the order of the ratio of the largest
dimension of the system to its formal Compton length.  To use our earlier
illustration, the bound on the information capacity of a compact disk is
now about $10^{40}$ bits, 28 orders tighter than the holographic
information bound.  But the universal bound is still many orders above
any foreseeable capacity.   Can one do better ?  One situation where this is
possible is for systems which are extensive in the thermodynamic sense. 
For these Gour~\cite{gour} has shown that $\mathcal{S}<(ER/\hbar
c)^{3/4}$ up to a numerical coefficient dependent on the number of
species.   For composite systems ($ER/\hbar c\gg 1$) Gour's bound is
tighter than the universal one, and while the former's scope is more
limited, it is nonetheless useful.  Other tight bounds for restricted
situations are sure to exist.   

\section{\label{sec:flow}     Bounds on information flow}

Information theory has perennially been very much a theory of
communication channel capacity.  Can one use black holes to obtain new
insights into natural limitations on information flow  rate ?  Indeed one
can, though in contrast to the case of the entropy bounds discussed earlier,
the results on information flow were actually known in some form from
early work on quantum communication channels. 

As before, we shall make use of the GSL.  The first thing to notice is that
the power in Hawking radiation of a Schwarzschild black hole can be
written as, c.f. Eq.~(\ref{flux}),
\begin{equation}  P_{BH}={{c^2\bar\Gamma\mathcal{N}}\hbar\over
15,360 \pi M^2}
\label{power}
\end{equation}
 According to out comments in connection with Eq.~(\ref{netchange}), the
entropy outflow $\dot S_{r}$ is $\nu P_{BH}/T_{{BH}}$.  We can thus
write
\begin{equation}
\dot S_{r}=\left({\pi\nu^{2}\bar\Gamma\mathcal{N}P_{PH}\over
240\hbar}\right)^{1/2}
 \label{Sdot}
\end{equation}
 (In Ref.~\cite{erice} the numerical coefficient under the radical is a factor
of 2 greater than here because there we consider photons as two separate
helicity species.)

This is our key formula.    Notably the entropy outflow versus power
relation here is not that of a typical 3-D hot body, for which $\dot
S_{r}\propto P^{{3/4}}$.  In terms of its radiative properties, a black hole
is more like a body thermally radiating in 1-D for which $\dot
S_{r}\propto P^{{1/2}}$~\cite{bek_mayo,erice}.  This observation leads us
right into the subject of communication channels which often are just 1-D
conduits of radiation.

We define a communication channel $\mathcal{C}$ very generally:  a
collection of modes for quantum particles, massless or massive, possibly
of various species, bosonic or fermionic.  The modes may span a narrow or
a broad range of frequencies, as well as a range of directions, but all like
modes differing only in the time at which they are emitted must be
included provided they all move from source to receiver.    Signals are
conveyed by populating the modes with quanta in accordance with a
probability distribution for the various quantum states (density matrices). 
We know that von Neumann's entropy is an upper bound on the
information conveyed~\cite{nielsen}.   When no record of the actual
choice of state of the fields in the channel is kept, $\mathcal{C}$ carries
only entropy, and we wish to set a bound on it.   
    
Our restriction that the whole time sequence of modes with given quantum
numbers be included allows us to speak of steady state channel power $P$
and von Neumann entropy flux rate $\dot \mathcal{S}(P)$ for
$\mathcal{C}$.     To get our result we imagine  directing the channel
upon a Schwarzschild black hole in such a way that the channel's energy
all goes down the black hole.  To prevent part of the incident radiation
from being scattered out of the hole, we require that the channel only
include suitable modes.  For example, spherical-like modes (perhaps so
shaped by a mirror system) are acceptable so long as the wavelengths
spectrum has an upper cutoff
$\lambda_c$ substantially smaller than the hole's scale $M$.    So are
wavepacket modes with small spread of direction and transversal
dimensions well under $M$, which means they also have a long
wavelength cutoff $\lambda_{c}$ considerably shorter than $M$.   We
pick the black hole scale $M=\xi\lambda_c$ with $\xi$ of order a few, so
that virtually all the channel's energy goes into the hole (see
Fig.~5).   For any given
$\mathcal{C}$ it is convenient to define a characteristic black hole power,
c.f. Eq.~(\ref{power}), 
\begin{equation} P_{c }\equiv
{{c^2\bar\Gamma\mathcal{N}}\hbar\over 15,360 \pi \lambda_{c}^2}
\approx 10^{{-4}}c^{2}\hbar\lambda_{c}{}^{-2}.
\label{condition}
\end{equation} For example, for optical wavelengths, $P_c\sim 1/30$ erg
s$^{-1}$.  We must emphasise that if the signal carriers are massive
particles, power here must include the rest energy flux.

Now $(P-P_{BH})c^{-2}$, the net rate of gain of black hole mass, causes a
gain in black hole entropy at a rate $\dot S_{{BH}} =(P-P_{BH})/T_{{BH}}$
(both of these quantities might be negative).  In addition, the emitted black
hole power is accompanied by radiation entropy rate $\dot S_{{BH}} =\nu
P_{BH}/T_{{BH}}$. The (obviously positive) sum of these two entropy
contributions, with the substitution $T_{BH}\rightarrow \hbar c/8\pi M$,
provides, by the GSL, a upper bound on $\dot \mathcal{S}(P)$, from
which we infer one on communication rate $\dot\mathcal{I}$ (always
expressed in bits s$^{-1}$):
\begin{equation}
\dot \mathcal{I}(P)<{8\pi\lambda_c \over \hbar c}\left[\xi P+{\nu-1
\over
\xi}P_c\right]\log_{2} e
\label{sum}
\end{equation} The bound is smallest for $\xi=[(\nu-1)P_{c}/P]^{1/2}$, at
which value the two terms in bound (\ref{sum}) are equal.  This
optimisation makes sense only if $\xi$ comes out of order a few at least,
that is for $P\ll P_{c}$.  In this case
\begin{equation}
\dot\mathcal{I}(P)<\left({\pi(\nu-1)\bar\Gamma\mathcal{N}P\over
60\hbar}\right)^{1/2} \log_{2}e
\label{ibound}
\end{equation}

\vspace{0.2in}
\includegraphics[height=2.0in]{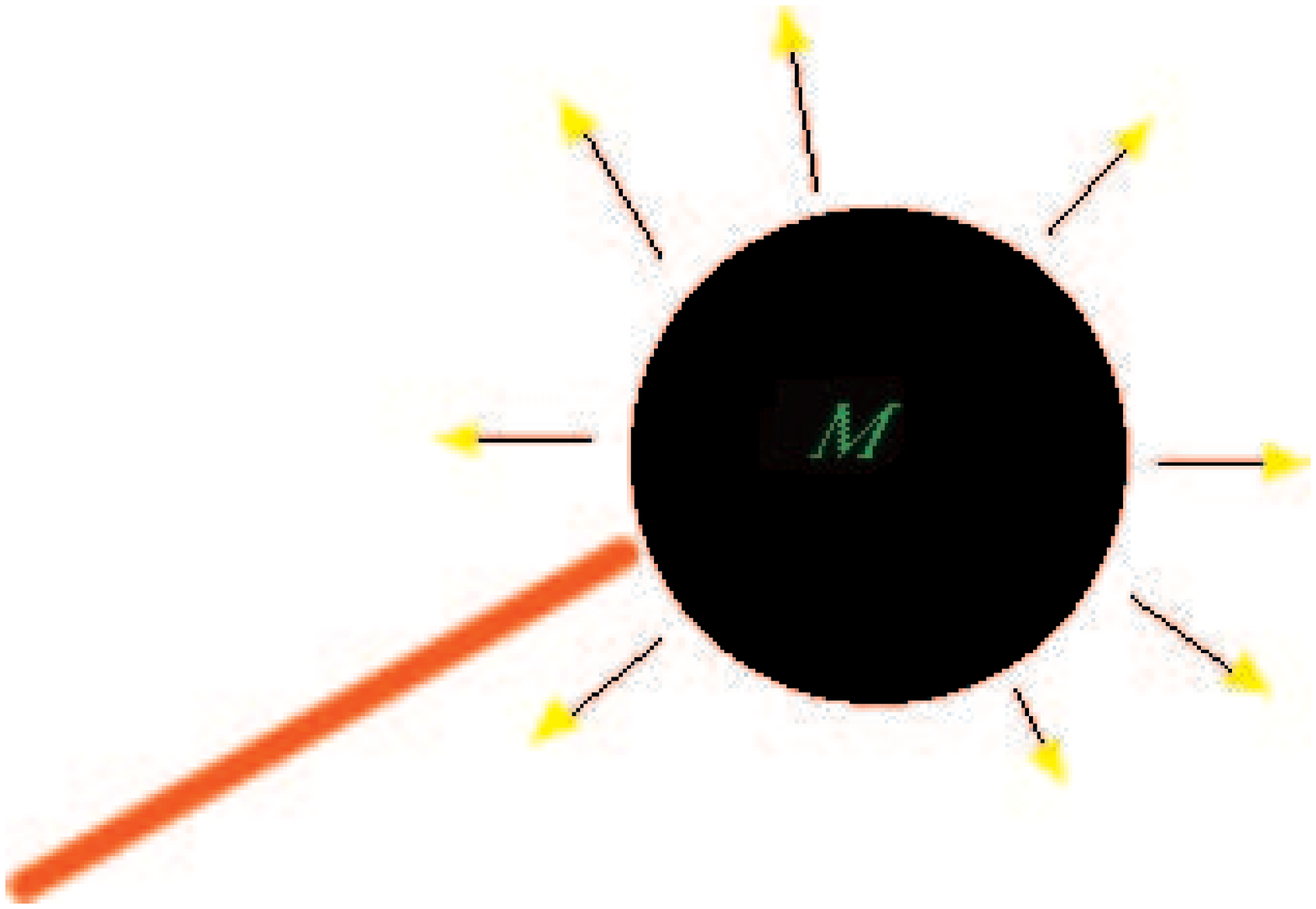}\\
{\bf Fig.~5: A communication channel is directed onto a Schwarzschild
black hole whose scale $M$ is chosen large compared to the longest
wavelength in the channel (which  determines its width).  Meanwhile the
hole radiates \emph{a la} Hawking.\hfill}
\vspace{0.2in}

For $P>P_c/10$, the optimal $\xi$ is no longer in a safe range for total
absorption.  For all such channels we simply fix $\xi=M/\lambda_c$ at
some large value (we shall take $\xi=10$ for illustration), and use  the full
bound (\ref{sum}).   If in addition  $P> P_c$, then by Eqs.~(\ref{power}) 
and (\ref{condition})  the second term in Eq.~(\ref{sum}) becomes
negligible.  Thus for  $P> P_c$
\begin{equation}
\dot \mathcal{I}(P)<{8\pi\xi\lambda_c P\over \hbar c}\log_{2} e.
\label{linear}
\end{equation} This bound is formally independent of the black hole
characteristic parameters $\bar\Gamma$ and $\nu$ and even of
$\mathcal{N}$.  

An immediate corollary of bound (\ref{linear}) is sometimes useful. 
Because of the cutoff $\lambda_c$, any information packet in the channel
cannot be of shorter duration than $\sim \lambda_c/c$.  Thus the total
energy $E$ of such packet must exceed $P\lambda_c/c$, so that 
\begin{equation}
\dot \mathcal{I} < {8\pi\xi  E\over \hbar }\log_{2} e.
\end{equation} This is Bremermann's bound~\cite{bremer} (usually
stated with a smaller coefficient).  It has been obtained in a variety of
ways, including derivation from the universal entropy
bound~\cite{bek81,bek88,bek_schiff}.  While independent of many details
and good for first orientation, it is rather lax.

What  to compare bounds (\ref{sum})--(\ref{linear}) with ?     Let us
restrict attention to channels employing massless particles for signalling. 
These include phonons in solids as well as photons in an optical fibre.  If
the channel has no short wavelength cutoff, then on dimensional grounds
an information \emph{packet} should have a maximum information
depending on its total energy $E$, its duration $\tau$ and the cutoff
$\lambda_c$.   No other variables seem relevant. Let us write
\begin{equation}
\mathcal{I}_{\rm max}=F\Big({\lambda_c^2 E^2\over
c^2\hbar^2},{E\tau\over
\hbar}\Big)
\label{Imax}
\end{equation} where $F$ is  a positive function of the indicated
dimensionless variables (no other independent ones can be formed from 
$E$, $\tau$ and $\lambda_c$).    If the packet's duration is long in some
sense, we can think of the flow of information as a  steady state with flow
rate $\dot\mathcal{I} =\mathcal{I}/\tau$ and power $P=E/\tau$.  The
channel capacity $\dot\mathcal{I}_{\rm max}$ deriving from
(\ref{Imax})  should obviously depend only on $P$ and $\lambda_c$, but
not $E$ and
$\tau$ separately.  This is possible only if $F$ is homogeneous of degree
$1/2$ with respect to both its arguments, $x$ and $y$.  But the most
general such function can be written $F(x,y)=\surd y\, f(x/y)$ with $f(z)$
another positive function.  Thus
\begin{equation}
\dot\mathcal{I}_{\rm max}=(P/\hbar)^{1/2} f(\lambda_c^2 P/
c^2\hbar).
\label{Idotmax}
\end{equation} We are being cavalier in treating $\tau$ in these
proceedings as if it were dimensionless.  However, it is clear that
Eq.~(\ref{Idotmax}) is dimensionally correct.  It should be remarked that
$f(z)$ should be a monotonically increasing function because the longer
the cutoff $\lambda_c$, the more modes are employed in the information
transport for one and the same $P$.

Let us consider the limit $\lambda_c\rightarrow\infty$.  The capacity of
such a cutoff-free channel is known~\cite{pendry,bek_schiff}.  Regardless
of the dispersion relation obeyed by the waves
\begin{equation}
\dot\mathcal{I}_{\rm max}=(n\pi P/3\hbar)^{1/2}\, \log_{2} e,
\label{pendry}
\end{equation} where $n$ stands for an effective number of information
carrier species.  Comparing with Eq.~(\ref{Idotmax}) we conclude that
\begin{equation} f(\infty)=(n\pi/3)^{1/2}\, \log_{2} e.
\label{f(infty)}
\end{equation}
 Evidently Eq.~(\ref{pendry}) should be applicable for any $\lambda_c$
provided $P$ is sufficiently large.   By comparing with
Eq.~(\ref{condition}) it is clear that the case before us is the high power
case of the black hole derived bound (\ref{sum}).   Indeed, bound
(\ref{linear}) is found to exceed the capacity (\ref{pendry}) even for
$P$ somewhat below $P_c$ and for $n$ as large as $\mathcal{N}$
($\mathcal{N}$ is evidently of the order of the maximum $n$ allowed).

Passing now to the low $P$ limit of capacity (\ref{Idotmax}),  we enter the
$P\ll P_c$ regime for which bound (\ref{ibound}) is relevant.  Comparing
the two we conclude that  
\begin{equation} f(0)<[\pi(\nu -1) \bar\Gamma\mathcal{N}/60]^{1/2}\,
\log_{2} e.
\label{f(0)}
\end{equation} This inequality together with Eq.~(\ref{f(infty)})  are 
generally consistent with the requirement of monotonic behaviour of
$f(z)$.  A problem could arise only if (\ref{f(0)}) is very close to saturation
\emph{and} $\mathcal{N}$ is larger than 20 (because we may always
chose
$n=1$, for example).  However, there is no reason to expect that bound
(\ref{ibound}) is anywhere close to equality, and the number of massless
species in nature is, after all, rather modest.  
 
\section{\label{sec:sum}Summary}

The GSL is the unifying thread of this review.  We have seen why it is
required for black hole containing systems.  It differs from the ordinary
second law in that black hole horizon area becomes a proxy for the 
entropy that has gone beyond the horizon.  This black hole entropy
complicates the usual question of consistency between unitary evolution
and the irreversibility required by the second law, thus engendering the
information paradox.  We have seen that there exist several resolutions to
this quandary in the spirit of unitary evolution, but that as a practical
matter, information is lost in black holes.

The GSL provides a good way to obtain  Susskind's form of the
holographic bound on the information storage capacity of a spatially finite
system.  We have seen that this simple bound can fail in certain
circumstances;  it must then be replaced by Bousso's covariant form of the
holographic principle.  This very general bound is one way to get at the
universal bound on the entropy of any finite weak self-gravity system,
which bound is much tighter than the primitive holographic bound.   We
have also supplied a GSL based derivation of the universal entropy
bound.  The holographic bounds are intimately related to 't Hooft's
holographic principle, the claim that the physics of a system is equivalent
to a theory restricted to its spatial boundary, but this principle, thus far
only partly established, is a stronger claim than those made by the various
entropy bounds. 

Finally, we have seen how to apply the GSL to the issue of quantum
communication channel capacity.  In contrast to the direct calculation  of
the entropy  for a precisely specified channel, as customary in information
theory, we have given an example of how to derive a bound on such
entropy for a more vaguely specified channel by applying the GSL to its
interaction with a black hole.  Variations of this example should be
illuminating. 

\section*{ Acknowledgments}   My former student  A. Mayo first taught
me some of the results in Sec.~\ref{sec:flow}. Research on this subject is
supported by the Israel Science Foundation established by the Israel
Academy of Sciences and Humanities.

\begin{quote}
\emph{Jacob Bekenstein} obtained his Ph. D. from Princeton University in
1972.  He was a postdoctoral fellow at the University of Texas at Austin,
and in 1974 moved to the new Ben Gurion University in Israel where he
became full professor in 1978 and the Arnow Professor of Astrophysics in
1983.  In 1990 he moved to the Hebrew University of Jerusalem where he
has been since 1993 the Polak Professor of Theoretical Physics.  In 1997 he
was elected to the Israel Academy of Sciences and Humanities.  He is a
member of the International Astronomical Union and is now serving a
second term in the International Committee for General Relativity and
Gravitation.  His scientific interests include gravitational theory, black
hole physics, relativistic magnetohydrodynamics, galactic dynamics, and
the physical aspects of information theory.
\end{quote}


\begin{thebibliography}{99}
\bibitem{michell}Michell, J., 1784, Phil. Trans. R. Soc. (London), {\bf 74},
35
\bibitem{laplace}Laplace, P. S., 1796, {\it Exposition du Systeme du
Monde} (Paris), Vol. II, p. 305; English edition: 1809, {\it The System of the
World} (London: W. Flint)
\bibitem{wheeler}Ruffini R. and Wheeler J. A., 1971, Physics Today, {\bf
24}, no. 12, 30
\bibitem{NC}Bekenstein J.D., 1972,  Lett. Nuovo Cim., {\bf 4}, 737
\bibitem{PRD1}Bekenstein J.D., 1973, Phys. Rev. D, {\bf 7}, 2333
\bibitem{hawking1}Hawking, S.W., 1975, Commun. Math. Phys.,  {\bf
43}, 199
\bibitem{PRD2}Bekenstein, J.D., 1974, Phys. Rev. D, {\bf 9}, 3292.
\bibitem{SciAm}Bekenstein, J.D., 2003, Scientific American, {289}, no. 2,
58-65
\bibitem{MTW}Misner  C. W.,  Thorne K. S. and  Wheeler J. A., 1973,  {\it
Gravitation\/}  (San Francisco: Freeman)
\bibitem{penrose}Penrose R. and  Floyd R. M., 1971, Nature,  {\bf 229},
177
\bibitem{christodoulou}Christodoulou D., 1970, Phys. Rev. Lett.,  {\bf 25},
1596 
\bibitem{hawking}Hawking S.W., 1971, Phys. Rev. Lett.,  {\bf 26}, 1344
\bibitem{PT}For a recollection of the early days of black hole
thermodynamics see Bekenstein J. D., 1980, Physics Today,  {\bf 33}, no. 1,
24 
\bibitem{gour_mayo}Gour, G. and Mayo, A.M., 2001, Phys. Rev. D, {\bf
63} 064005.
\bibitem{limits}Bekenstein J.D., 2001, Stud. Hist. Phil. Mod. Phys.,  {\bf
32}, 511
\bibitem{brazil}Bekenstein J.D.,  2000, Black holes: classical properties,
thermodynamics and heuristic quantisation. In {\it Cosmology and
Gravitation\/}, edited by M. Novello (Paris: Atlantisciences) pp. 1-85.
\bibitem{BKLS}Bombelli, L. , Koul, R. Lee, J.  and
 Sorkin, R., 1986, Phys. Rev. D. {\bf 34}, 373
\bibitem{sred}Srednicki, M., 1993, Phys. Rev. D. {\bf 71}, 66
\bibitem{TZ}Thorne, K. S. and Zurek, W. H., 1985, Phys. Rev. Letters,
{\bf   54}, 2171 
\bibitem{thooft2}'t Hooft, G., 1985, Nucl. Phys. B, {\bf 256}, 726
\bibitem{strom_wafa}Strominger, A. and Wafa, C., 1996, Phys. Lett. B,
{\bf 379}, 99
\bibitem{malda_strom}Maldacena, J., Strominger, A. and Witten, E., 1997,
J. High Energy Phys.,  {\bf 9712}, 002
\bibitem{carlip}Carlip, S., 1999, Class.  Quant. Grav. {\bf 16}, 3327 
\bibitem{solo}Solodukhin, S.N., 1999, Phys. Lett. B {\bf 454}, 213
\bibitem{ash}Ashtekar, A., Corichi, A., Baez, J. and Krasnov, K., 1998,
Phys. Rev. Lett. {\bf 80}, 904
\bibitem{mukh}Bekenstein, J.D. and Mukhanov, V. F, 1995, Phys. Lett. B
{\bf 360},  7
\bibitem{MG8}Bekenstein, J.D. 1999, Quantum black holes as atoms.  In
{\it Proceedings of the Eight Marcel Grossmann  Meeting on  General
Relativity},  edited by T. Piran and R. Ruffini,  (Singapore: World
Scientific), pp. 92-111
\bibitem{hawking2}Hawking S.W.,  1976, Phys. Rev. D, {\bf 14 },  2460
\bibitem{bek94}Bekenstein J.D., 1994, Phys. Rev. D, {\bf 49}, 912
\bibitem{bound}Bekenstein J.D., 1981, Phys. Rev. D, {\bf 23}, 287
\bibitem{susskind}Susskind L., 1995, J. Math. Phys., {\bf 36}, 6377
\bibitem{catalyse}Bekenstein J.D., 2000, Phys. Letters, {\bf 481}, 339;  2002,
Holographic bound from second law.  In {\it Proceedings of the Ninth
Marcel Grossmann  Meeting on  General Relativity},  edited by V. G.
Gurzadyan, R. Jantzen and R. Ruffini,  (Singapore: World Scientific), pp.
553-559
\bibitem{bousso00}Bousso, R., 1999, J. High Energy Phys., {\bf 9907}, 004
\bibitem{boussorev}Bousso, R., 2002, The holographic principle, Rev.
Mod. Phys. {\bf 74}, 825-874
\bibitem{loewe}Loewe, D., 1999, J. High Energy Phys., {\bf 9910}, 026.
\bibitem{strom}A. Strominger and D. Thompson, 2003, A quantum
Bousso bound.  LANL preprint hep--th/0303067
\bibitem{thooft}'t Hooft, G.,  1993, Dimensional reduction in quantum
gravity.  In  {\it Salam--festschrifft\/}, edited by A. Aly, J. Ellis,  and  S.
Randjbar--Daemi  (Singapore: World Scientific),  LANL preprint
gr--qc/9310026
\bibitem{maldacena}Maldacena, J., 1998, Adv. Theor. Math. Phys. {\bf 2},
231
\bibitem{witten}Witten, E., 1998, Adv. Theor. Math. Phys. {\bf 2}, 253
\bibitem{gibb}Gibbons, G. and Hawking, S. W., 1977, Phys. Rev. D {\bf
15}, 2738
\bibitem{davies}Davies, P.C.W. and Davis, T. M., 2002, Found. Phys. {\bf
32}, 1877
\bibitem{buoy}Unruh, W.G.  and Wald, R.M., 1982, Phys. Rev. D, {\bf 25},
942; 1983, Phys. Rev. D, {\bf 27}, 2271.  Bekenstein, J.D., 1982, Phys. Rev.
D, {\bf 26}, 950; (1983) Phys. Rev. D, {\bf 27}, 2262; (1994) Phys. Rev. D,
{\bf 49}, 1912; (1999) {\it Phys. Rev. \/}D, {\bf 60}, 124010. 
\bibitem{erice}Bekenstein, J.D., 2002, Quantum information and quantum
black holes.  In {\it Advances in the Interplay Between Quantum and
Gravity Physics}, edited by P. Bergmann and V. de Sabbata (Dordrecht:
Kluwer), pp 1-85
\bibitem{page2}Page, D.N., 1983, {\it Phys. Rev. Letters\/}, {\bf 50}, 1013
\bibitem{page1}Page, D.N., 1976, Phys. Rev. D, {\bf 13}, 198
\bibitem{bousso03}Bousso, R., 2003, Phys. Rev. Lett. {\bf 90}, 121302
\bibitem{BFM}Bousso, R., Flanagan, E. E. and Marolf, D., 2003,  Phys. Rev.
D, {\bf 68}, 064001
\bibitem{husain}Husain, V, 2003, Probing entropy bounds with scalar
field spacetimes, LANL preprint gr-qc/0307070
\bibitem{bek84}Bekenstein J.D., 1984, Phys. Rev. D, {\bf 30}, 1669;
Bekenstein, J.D. and Schiffer, M.,  1989, Phys. Rev. D, {\bf 39}, 1109
\bibitem{bek_schiff}Bekenstein, J.D. and Schiffer, M., 1990, Int. J. Mod.
Physics C, {\bf 1}, 355
\bibitem{gour}Gour, G., 2003, Phys. Rev. D, {\bf 67}, 127501
\bibitem{bek_mayo}Bekenstein, J.D. and Mayo, A.E., 2001, Gen. Rel. Grav.
{\bf 33}, 2095-2099
\bibitem{nielsen}Nielsen, M. A. and Chuang, I. L., 2000,  {\it Quantum
Computation and Quantum Information\/}  (Cambridge: Cambridge
University Press)
\bibitem{bremer}Bremermann, H.J., 1967, {\it Proceedings of Fifth
Berkeley Symposium on Mathematical Statistics and Probability\/},
edited by L. M. LeCam, and J. Neyman,  (Berkeley: Univ. of California
Press)
\bibitem{bek81}Bekenstein J.D., 1981, Phys. Rev. Letters, {\bf 46}, 623.
\bibitem{bek88}Bekenstein J.D., 1988,   Physical Review A, {\bf 37}, 3437
\bibitem{pendry}Pendry, J.B., 1983, J. Phys. A, {\bf 16}, 2161


\end{thebibliography}
\end{document}